\documentclass[12pt,a4paper]{iopart}

\usepackage[english]{babel}
\usepackage{iopams}
\usepackage{ae} 
\usepackage{subfigure}
\usepackage{color}
\usepackage{ifthen}
\usepackage{bm}

\usepackage[pdftex]{graphicx}
\usepackage[pdftex]{hyperref} 
\hypersetup{colorlinks=true,
  pdfstartview=FitV,
  linkcolor=blue,
  citecolor=blue,
  urlcolor=blue,
  pdfauthor={HORVATH Laszlo, laszlo.horvath@ukaea.uk},
  pdfsubject={IAEA TM on EPs 2015},
  pdftitle={Horvath IAEA TM on EPs 2015 paper}
  }
\DeclareGraphicsExtensions{.pdf}

\newcommand{\eqref}[1]{(\ref{#1})}

\usepackage{perpage} 
\MakePerPage{footnote}

\graphicspath{{figures/}}

\begin{document}

\title{Experimental investigation of the radial structure of energetic particle driven modes}
\author{L. Horv\'{a}th$^{1,2}$, G. Papp$^{3,4}$, Ph. Lauber$^3$, G. Por$^1$, A. Gude$^3$, \\
  V. Igochine$^3$, B. Geiger$^3$, M. Maraschek$^3$, L. Guimarais$^5$,\\
  V. Nikolaeva$^{3,5,6}$, G. I. Pokol$^1$ and the ASDEX Upgrade Team$^3$}

\address{${}^1${}Institute of Nuclear Techniques, Budapest University of Technology and Economics, Pf 91, H-1521 Budapest, Hungary\\
  ${}^2${}York Plasma Institute, Department of Physics, University of York, Heslington, York, YO10 5DD, UK\\
  ${}^3${}Max-Planck-Institute for Plasma Physics, D-85748 Garching, Germany\\
  ${}^4${}Max-Planck/Princeton-Center for Plasma Physics\\
  ${}^5${}Instituto de Plasmas e Fus\~ao Nuclear, Instituto Superior T\'ecnico, Universidade de Lisboa, Portugal\\
  ${}^6${}Physik-Department E28, Technische Universit\"at M\"unchen, 85748 Garching, Germany}

\ead{laszlo.horvath@ukaea.uk}

\begin{abstract}
Alfv\'{e}n eigenmodes (AEs) and energetic particle modes (EPMs) are often excited by energetic particles (EPs) in tokamak plasmas.
One of the main open questions concerning EP driven instabilities is the non-linear evolution of the mode structure.
The aim of the present paper is to investigate the properties of beta-induced AEs (BAEs) and EP driven geodesic acoustic modes (EGAMs) observed in the ramp-up phase of off-axis NBI heated ASDEX Upgrade (AUG) discharges.
This paper focuses on the changes in the mode structure of BAEs/EGAMs during the non-linear chirping phase.
Our investigation has shown that in case of the observed down-chirping BAEs the changes in the radial structure are smaller than the uncertainty of our measurement.
This behaviour is most probably the consequence of that BAEs are normal modes, thus their radial structure strongly depends on the background plasma parameters rather than on the EP distribution.
In the case of rapidly upward chirping EGAMs the analysis consistently shows shrinkage of the mode structure.
The proposed explanation is that the resonance in the velocity space moves towards more passing particles which have narrower orbit widths.


\end{abstract}

\pacs{52.25.Xz, 52.55.Fa, 52.35.Bj, 52.35.Mw, 52.55.Pi}

\vspace{2pc}
\noindent{\it Keywords}: tokamak, energetic ions, Alfv\'{e}n eigenmodes, wave-particle interaction, time-frequency analysis
%
%
\maketitle
%

\section{Introduction}

Supra-thermal energetic particles (EPs) in tokamak plasmas can excite various instabilities, which can lead to an enhanced transport of fast particles~\cite{heidbrink08basic}.
The understanding of these instabilities is crucial, because the most important transport process of EPs in the plasma core is their interaction with global plasma modes~\cite{lauber13super}.
There are two main types of fast particle - wave interactions.
One is when energetic particles drive weakly damped Alfv\'{e}n eigenmodes unstable. These modes are typically associated with a gap in the Alfv\'{e}n continuum in order to avoid continuum damping.
The other type are the so called energetic particle modes (EPMs)~\cite{heidbrink08basic}, which usually appear when the EP pressure is comparable to the thermal pressure and the drive can overcome the continuum damping \cite{chen16physics}.

Several types of non-linear behaviour of the mode amplitude and frequency are observed on present-day tokamaks.
The type of behaviour is expected to significantly influence the impact of the instabilities on the fast particle transport, thus their thorough understanding is essential.
In this paper modes with bursting amplitude and rapidly changing mode frequency -- called {\it chirping modes} -- are investigated experimentally.
The detailed analysis of chirping beta induced Alfv\'{e}n eigenmodes (BAEs)~\cite{turnbull93global} and EP-driven geodesic acoustic modes (EGAMs)~\cite{fu08energetic} observed in the ramp-up phase of off-axis neutral beam injection (NBI) heated plasmas in ASDEX Upgrade (AUG)~\cite{lauber13offaxis} is presented.

The BAE gap is introduced by the coupling between the compressible ion acoustic branch and the shear Alfv\'{e}n continuum via geodesic curvature~\cite{turnbull93global}.
BAE is an electromagnetic mode with non-zero toroidal mode number.
The mode frequency can  be estimated with the simplified (this does not contain the $1/q^2$ correction~\cite{lauber13super}) dispersion relation $\omega^2_{\mathrm{BAE}} \approx {v^2_{\mathrm{th}}} (7/4 + T_e/T_i) /{R^2_0}$~\cite{mikhailovsky73drift,tang80kinetic,lauber13nbi}, where $v_{\mathrm{th}}$ is the ion thermal velocity, $R_0$ is the major radius of the magnetic axis, $T_e$ and $T_i$ are the electron and ion temperatures, respectively.
BAEs can be driven unstable by the radial gradient in the EP distribution function~\cite{nguyen09excitation}.
Thus, their presence enhances the radial transport of fast ions.
However, their stability in ITER is not yet clear.
Simulations have shown that in the ITER inductive baseline scenario the radial gradients in the EP distribution will not be sufficiently high to destabilize BAEs, but the difference is marginal and the uncertainty of these results is high~\cite{pinches15energetic}.
Fast particle driven BAEs have been previously observed in many tokamaks.
NBI ion driven chirping BAEs have been reported by DIII-D~\cite{heidbrink93observation, heidbrink99what}.
In Tore-Supra BAEs driven by ion cyclotron resonance heating (ICRH) generated ions have been investigated~\cite{nguyen09excitation} and energetic electron driven BAEs have been observed in HL-2A~\cite{chen10beta}. However, in the latter two cases, the mode activity exhibits no frequency chirping.

The EGAM is an EPM which is driven unstable by the velocity space gradient in the EP distribution~\cite{fu08energetic}.
It is an electrostatic mode with zero toroidal mode number ($n = 0$).
The EGAM frequency is at about half of the GAM frequency in many cases~\cite{nazikian08intense}, but in general the frequency ratio can be different~\cite{girardo14relation}.
The isotropic distribution of fusion-born $\alpha$ particles does not affect EGAMs, but the non-isotropic NBI distribution can excite these modes.
Although the canonical angular momentum ($P_{\phi}$) of the particles is not effected by the EGAM, the change of the energy of passing particles due to EGAMs is accompanied not only by the change of the pitch-angle parameter ($\lambda$), but also by a small change of the radial coordinate~\cite{kolesnichenko04cooling}.
Furthermore, EGAMs can influence fast ion losses indirectly via mode-mode coupling with TAEs and it has also been found that EGAMs can enhance the turbulent transport, leading to degradation of the transport barrier~\cite{zarzoso13impact}.
The experimental observation of EGAMs has been reported by various devices.
Chirping EGAMs driven by counter-injected NBI ions have been observed in DIII-D~\cite{nazikian08intense} and 
ICRH ion driven chirping EGAMs have been detected in JET~\cite{boswell06observation, berk06explanation}.
Energetic electron driven GAMs have been reported by HL-2A~\cite{chen13egam, chen13observation}.

Radial structure analysis of BAEs has been recently carried out on AUG by 2D electron cyclotron emission imaging (ECEI)~\cite{lauber13nbi,classen11investigation}.
However, that work reported the radial movement of the modes caused by the evolving background plasma parameters on a longer time scale ($\sim100$ ms).
In contrast, the main goal of this work is to experimentally investigate the rapid changes in the radial structure of bursting EP-driven modes during the non-linear chirping phase, on a $\sim1$ ms time scale.
The investigation of these modes is essential in order to understand non-linear wave-particle interaction.
Due to diagnostic and data analysis complexities this task has never been accomplished before.
Some modes are expected to retain their radial structure, while others would be expected to change.
Even qualitative results can provide important information about the underlying physics and strengthen (or challenge) our present theoretical understanding.

The radial structure analysis presented in this paper shows that in case of the observed downward chirping BAEs the changes in the radial eigenfunction were smaller than the uncertainty of the measurement, while in case of rapidly upward chirping EGAMs our results indicate shrinkage of the mode structure.
An explanation of the experimental observations based on the simulated equilibrium fast ion distribution is proposed.

The paper is organized as follows.
In section~\ref{sec:setup} we present the diagnostic and analysis tools we used for the investigation.
The scenario and a list of the examined discharges are shown in section~\ref{sec:scenario}.
The experimental investigation of BAEs and EGAMs are presented in section~\ref{sec:bae} and~\ref{sec:egam}, respectively.
Section~\ref{sec:egam} also contains the theoretical explanation of the observed behaviour of EGAMs; followed by the summary and conclusions in section~\ref{sec:conc}.

\section{Measurement set-up and analysis principles}\label{sec:setup}

The frequency of the BAEs and EGAMs investigated in this paper is in the order of $50$~-~$100$~kHz.
Both the diagnostic tools and the data processing method were chosen to deal with such high frequency oscillations.
From the diagnostic point of view, fluctuation measurements are required which can measure either magnetic, density or  temperature fluctuations in the plasma with temporal resolution higher than $100$~kHz.
The possible candidates were the magnetic pick-up coils, reflectrometry, electron cyclotron emission (ECE) and soft X-ray (SXR) measurements.

\begin{figure}[htb!]\centering
  \includegraphics[width = 120mm]{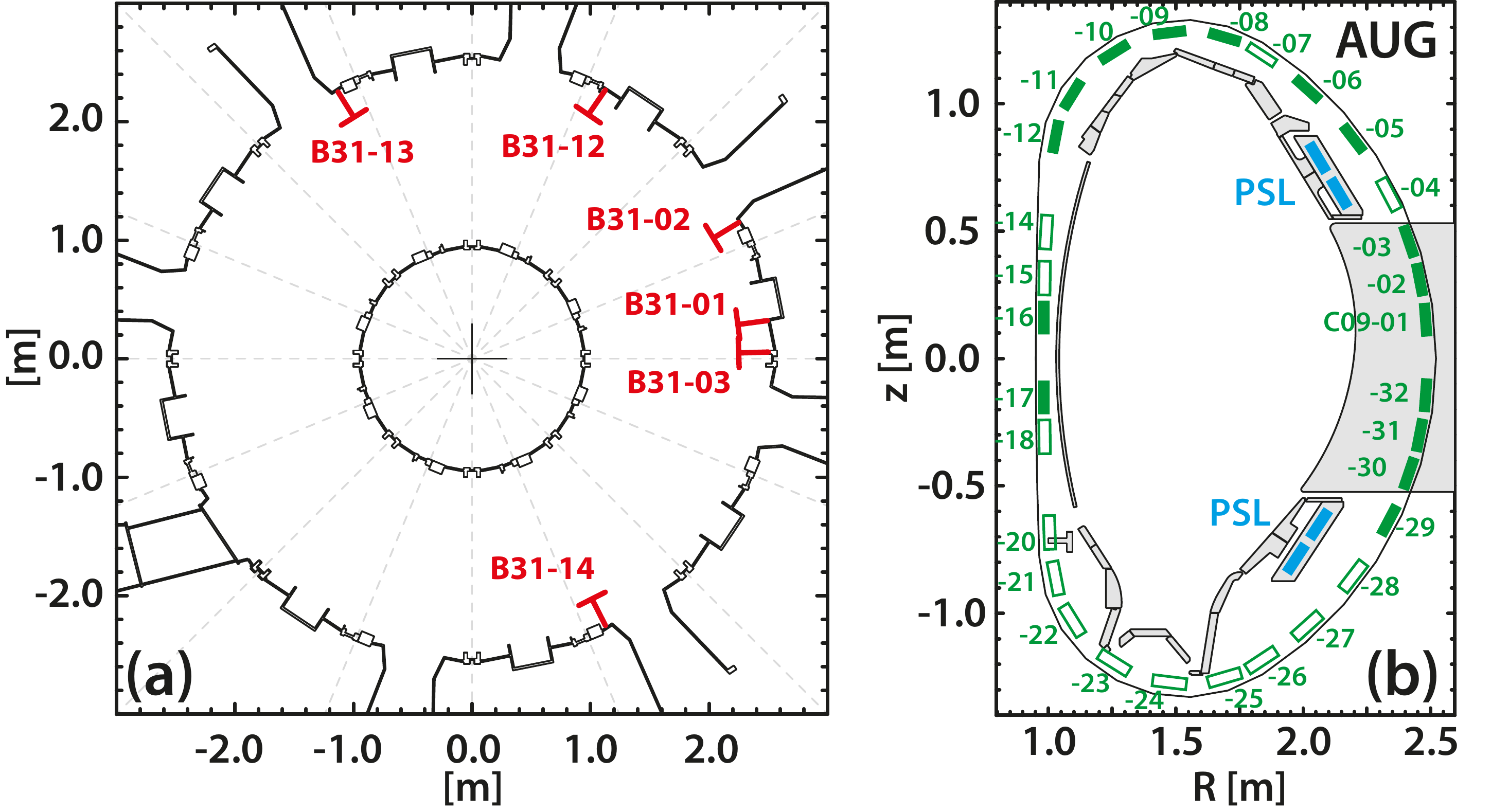}
  \caption{The position of magnetic probes used for the analysis on AUG. \textbf{(a)}~The six ballooning coils of the toroidal array are indicated with red ``T'' symbols in the top-down view of the tokamak. \textbf{(b)}~The 30 Mirnov coils indicated with green rectangles in a poloidal cross section of the device. Probes indicated with empty rectangles were ruled out from the analysis, because the investigated modes were not visible in their signal. The blue rectangles indicate the passive stabilizing loops (PSLs).}
  \label{fig:magnetic_probe_position}
\end{figure}

Magnetic measurements do not have radial resolution, thus these are not suitable for the analysis of the radial structure. Since several coils are distributed around the vessel, as is shown in figure~\ref{fig:magnetic_probe_position}, these were used to examine poloidal and toroidal mode numbers~\cite{horvath15reducing}.
Furthermore, magnetic spectrograms are excellent to follow the time evolution of the mode frequencies.
Note that in principle EGAMs should only be detectable with density fluctuation measurements because the EGAM is an electrostatic mode. However, due to sideband coupling it is clearly visible on magnetic fluctuation measurements as well~\cite{wahlberg08geodesic}.

The reflectrometry diagnostic can measure density fluctuations with high temporal resolution.
Due to the low spatial resolution, reflectrometry could not be used for the radial structure analysis, because the modes were visible on 2 channels at most.
However, the information provided by the reflectrometry (in addition to the SXR measurements) was useful in the radial localization of the modes.
The ECE and the ECE imaging (ECEI)~\cite{classen10electron} diagnostics would be good candidates due to their high time- and spatial resolution, however BAEs and EGAMs were not visible by ECE and ECEI in the investigated discharges (bad signal to noise ratio).
Thus, ECE and ECEI were ruled out from the analysis.

The only fluctuation measurement which features good spatial resolution and well resolves the modes was the  SXR diagnostic.
Since the SXR diagnostic is a line-integrated measurement, it is not straightforward to reconstruct the mode structure, but many line-of-sights (LOSs) are available which makes it possible to qualitatively investigate the time evolution of the radial structure.
The energy spectrum of soft X-ray radiation consists of a continuum radiation resulting from Bremsstrahlung, recombination radiation and line radiation.
In a plasma the emitted intensity of Bremsstrahlung radiation can be estimated by the following formula~\cite{huba09nrl}:
\begin{equation}\label{eq:sxr_intensity}
  P_{\mathrm{Br}} = 1.69\cdot10^{-32}\cdot n_e\sqrt{T_e}\sum \big( Z^2 N(Z) \big) ,
\end{equation}
where $n_e$ is the electron density in $1/\mathrm{cm}^{-3}$, $T_e$ is the electron temperature in eV and $Z$ is the charge number of the given ionization state.
The sum is executed on all ionization states and the radiation power is given in W/$\mathrm{cm}^{-3}$.
Equation~\eqref{eq:sxr_intensity} shows that the total radiated power density of Bremsstrahlung emitted by a deuterium plasma is proportional to the square of the electron density and the square root of the electron temperature.
However, the {\it detected} SXR emissivity is not proportional to $\sqrt{T_e}$, the $T_e$ dependence is modified by the beryllium filter installed on the pinhole detector.
From the analysis point of view, it is sufficient to know that the emissivity is a monotonic function of $T_e$ in the temperature range (below $10$ keV) relevant for the discharges investigated in this paper.

\begin{figure}[htb!]\centering
  \includegraphics[width = 50mm]{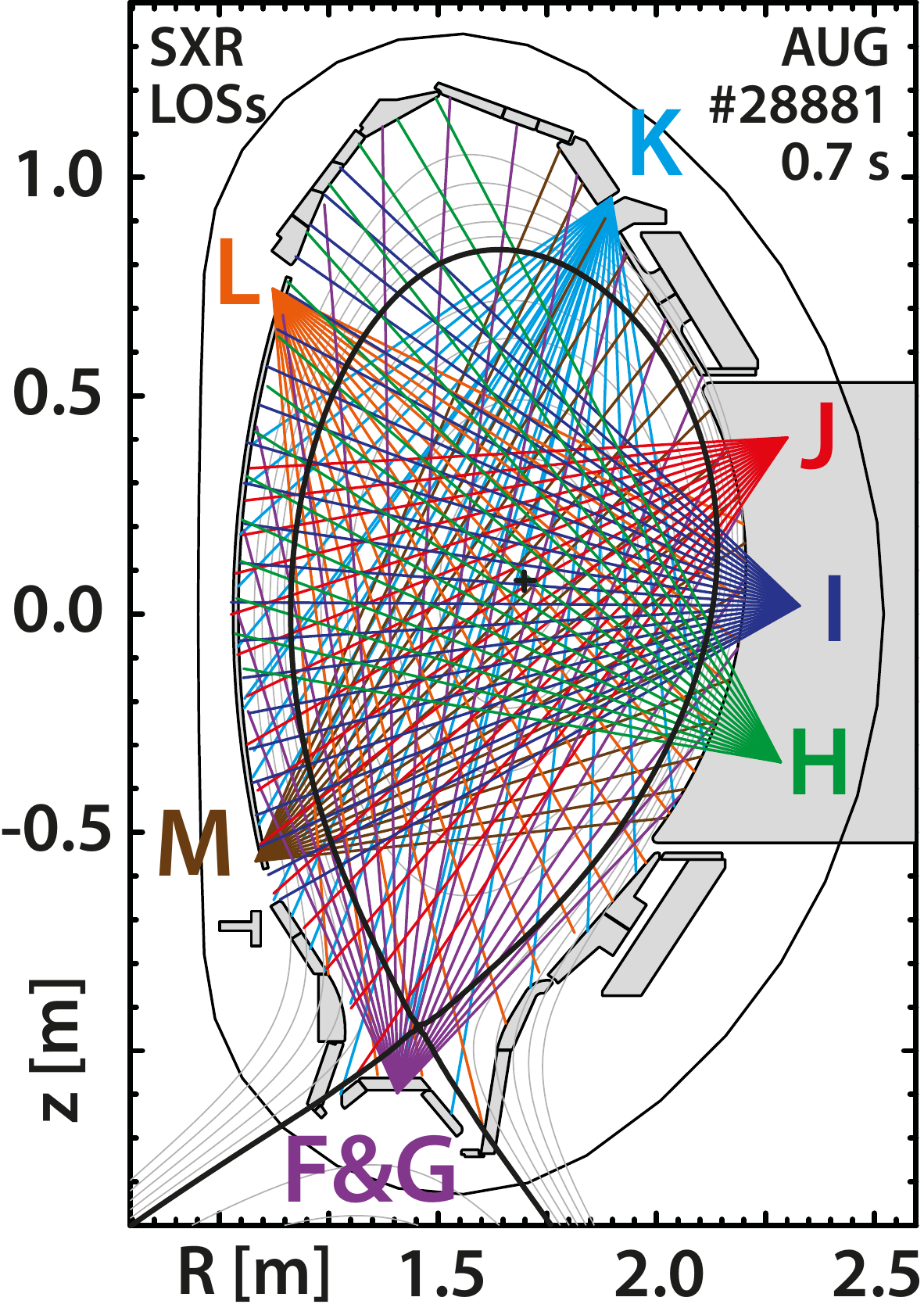}
  \caption{The line-of-sights (LOSs) of the used soft X-ray channels in the poloidal cross-section of AUG. Different colours indicate the different cameras.}
  \label{fig:sxr_position_all}
\end{figure}

During the analysis the signal of the majority of the LOSs shown in figure~\ref{fig:sxr_position_all} were examined.
In order to observe the changes in the radial structure of the modes, the oscillation amplitude has to be reconstructed for each SXR LOSs.
The oscillation caused by the observed mode was modelled as a frequency and amplitude modulated harmonic wave.
The noise of the measured signal was dominated by the fluctuations of the soft X-ray emission coming from the background plasma, which was modelled as a Gaussian, additive, white noise.
The noise of the amplifier chain was negligible.

The time evolution of the oscillation amplitude of the chirping modes was reconstructed using an advanced amplitude reconstruction method~\cite{horvath16time, horvath15analysis}:
\begin{equation}\label{eq:stft_linear_4}
  a(u) = \frac{| S f_{\mathrm{T}, u}(u, \xi) |}{\sqrt{2\sigma_t\sqrt{\pi}}} \sqrt[4]{1 + \phi''(u)^2\sigma_t^4} ,
\end{equation}
where $S f_{\mathrm{T}, u}(u, \xi)$ is the Short Time Fourier Transform (STFT)~\cite{mallat08wavelet, gabor1946analysis} of the measured signal $f$ at the mode frequency at a given time point $u$,
$\sigma_t$ is the standard deviation of the Gabor atom~\cite{mallat08wavelet} applied to calculate the transform and $\phi''(u)$ denotes the first derivative of the frequency ($\phi'' \equiv \omega'$).
The mode frequency was traced by a local maximum searching algorithm~\cite{horvath14changes, papp11low} on the STFT spectrogram, which follows the mode evolution until the oscillation amplitude falls below the background noise level.
Since the application of STFT serves as a narrow band filter, the white noise approximation of the background noise only assumes that the spectral power of the noise was uniform in a narrow frequency range ($\Delta f\sim10$~kHz).

Equation~\eqref{eq:stft_linear_4} is a linear chirp approximation which is a good estimate when the frequency and the amplitude of the chirping wave changes linearly on the time scale of~$\sigma_t$.
The validity of this approximation in a given case can be tested by reconstructing the amplitude using different time-frequency resolutions for STFT and applying the correction assuming linear chirp.
If the correction completely eliminates the biasing of the reconstructed amplitude by the different time-frequency resolutions, the linear chirp assumption is proven~\cite{horvath16time, horvath15analysis}.
This formula differs from the zeroth order approximation in the $\sqrt[4]{1 + \phi''(u)^2\sigma_t^4}$ correction factor.
This correction depends on the derivative of the frequency and the width of the applied Gabor atom, which determines the time-frequency resolution of STFT.
It is clear that the correction factor is lower if the window width is smaller, i.e when the time resolution of the transform is better.
However, the window width cannot be arbitrary small, because the better the time-resolution the worse the frequency resolution.
Therefore, a trade-off between the time- and frequency resolution is necessary.

The oscillation amplitudes shown in section~\ref{sec:bae} and~\ref{sec:egam} were reconstructed by using~\eqref{eq:stft_linear_4} and the error bars were derived from the standard deviation of the background noise by taking into account the error propagation in the reconstruction formula~\cite{horvath16time, horvath15analysis}.

\section{Scenario}\label{sec:scenario}

The plasma scenario on AUG in which bursting BAEs and EGAMs were observed needs low density ($\sim2\cdot10^{19}\ \mathrm{m}{}^{-3}$) operation and strong off-axis NBI heating.
The strong NBI heating and the low density are important to provide large enough drive for the modes by keeping the ratio of the fast ion pressure to the thermal pressure as high as possible.
On the other hand the main damping mechanism of BAEs and EGAMs is ion Landau damping, which scales with the background ion temperature and density.
Thus, it is easier to excite these modes with off-axis NBI, since off-axis drive means lower local ion temperature i.e. lower Landau damping.
Furthermore, the elevated $q$ profile in the ramp-up also leads to lower Landau damping.
On AUG, even in low density operation, the damping in the flat-top is not sufficiently small to excite these modes with NBI.
However, in the ramp-up phase these modes are routinely driven unstable~\cite{lauber13nbi, classen11investigation}.
A series of discharges with strong Alfv\'{e}nic activity were investigated.
The list of analysed discharges with the approximate time and frequency intervals where BAEs and EGAMs were observed is presented in table~\ref{tab:shots}.
\begin{table}[htb!]
\footnotesize
  \begin{center}
    \begin{tabular}{|c||c|c||c|c|}
\hline
    &\multicolumn{2}{c||}{BAEs}	&\multicolumn{2}{c|}{EGAMs}	\\
	&Time [s]	&Frequency [kHz]	&Time [s]	&Frequency [kHz]	\\
\hline\hline
\#25506	&0.45 - 0.60	&50 - 60	&$\times$	&$\times$\\
\#28881	&0.64 - 0.72	&70 - 80	&0.85 - 1.00	&60 - 80\\
\#28884	&0.62 - 0.72	&70 - 85	&$\times$	&$\times$\\
\#28885	&$\times$	&$\times$	&0.75 - 0.95	&70 - 85\\
\#30383	&$\times$	&$\times$	&0.25 - 0.60	&30 - 60\\
\#30946	&0.45 - 0.60	&40 - 70	&0.60 - 0.70	&40 - 60\\
\#30950	&0.45 - 0.55	&50 - 60	&0.70 - 0.80	&60 - 80\\
\#30951	&0.55 - 0.70	&70 - 90	&$\times$	&$\times$\\
\#30952	&0.40 - 0.45	&60 - 80	&0.50 - 0.70	&50 - 70\\
\#30953	&$\times$	&$\times$	&0.55 - 0.75	&40 - 60\\
\#31213	&$\times$	&$\times$	&0.65 - 1.00	&40 - 60\\
\#31214	&$\times$	&$\times$	&0.45 - 0.80	&60 - 80\\
\#31215	&$\times$	&$\times$	&0.45 - 0.75	&40 - 60\\
\#31216	&$\times$	&$\times$	&0.50 - 0.60	&50 - 70\\
\#31233	&$\times$	&$\times$	&0.80 - 1.00	&60 - 80\\
\#31234	&$\times$	&$\times$	&0.65 - 0.80	&40 - 50\\
\hline
    \end{tabular}
    \caption{\label{tab:shots} Investigated discharges with strong Alfv\'{e}n activity. The approximate time and frequency intervals where BAEs and EGAMs were observed are presented.}
  \end{center}
\end{table}

The radial structure analysis was carried out by using the SXR measurement system.
The change in the DC component of the SXR signals and the change of the background plasma parameters on the time scale of a chirp (typically $\sim1-10$ ms) were negligible, which indicates that any observed change in the radial structure of the modes was an effect of the evolving fast ion distribution function.
In general, the signal-to-noise ratio of the SXR measurements in terms of the observed mode amplitude was poor for the purpose of radial structure analysis.
In many cases, modes which were clearly visible on the magnetic spectrogram were well distinguishable on only one or two LOSs of SXR.
The strategy was to find cases where the mode is observable on at least 3 adjacent LOSs of a particular SXR camera.
LOSs of SXR cameras F, G, H, I and J (see figure~\ref{fig:sxr_position_all}) were investigated in the discharges listed in table~\ref{tab:shots} by calculating the spectrogram of their signal in the time-frequency range where the mode was visible on the magnetic spectrogram.
In total, about $1000$ SXR spectrograms were analysed in this process.
Finally, 3 cases for BAEs and 5 cases for EGAMs were selected, where the oscillation amplitude on the SXR signals was sufficiently high to identify them visually on the SXR spectrograms.

\section{Beta-induced Alfv\'{e}n eigenmodes}\label{sec:bae}

\begin{figure}[htb!]\centering
  \includegraphics[width = 80mm]{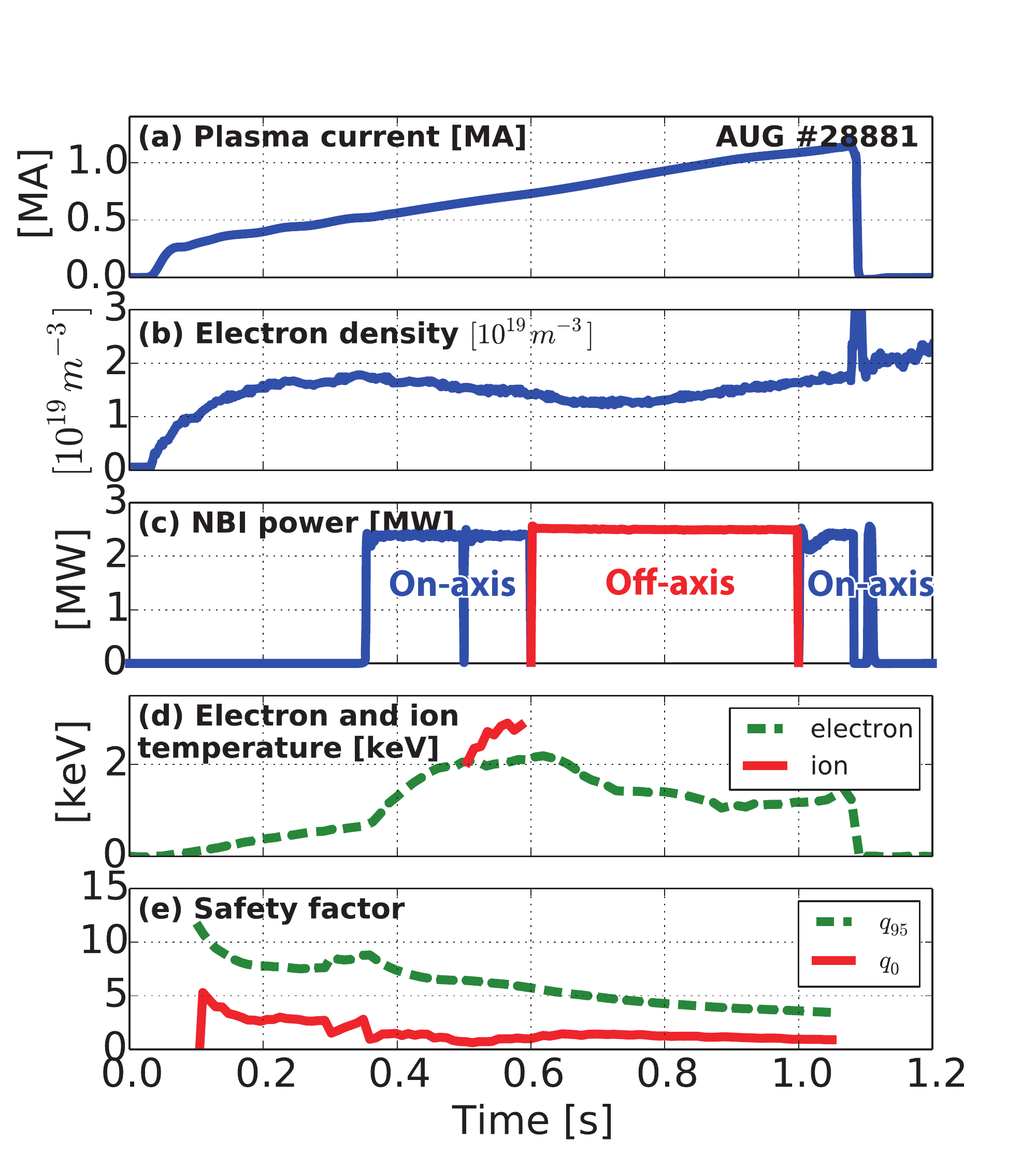}
  \caption{Main waveforms of discharge \#28881: \textbf{(a)}~plasma current, \textbf{(b)}~line-integrated core electron density, \textbf{(c)}~NBI power, \textbf{(d)} core electron temperature with green dashed line and core ion temperature with red solid line (the ion temperature measurement was available from $0.5$ s to $0.6$ s and \textbf{(e)}~safety factor $q$ at the magnetic axis with red solid line and $q_{95}$ with green dashed line.}
  \label{fig:28881}
\end{figure}

\begin{figure}[htb!]\centering
  \includegraphics[width = 145mm]{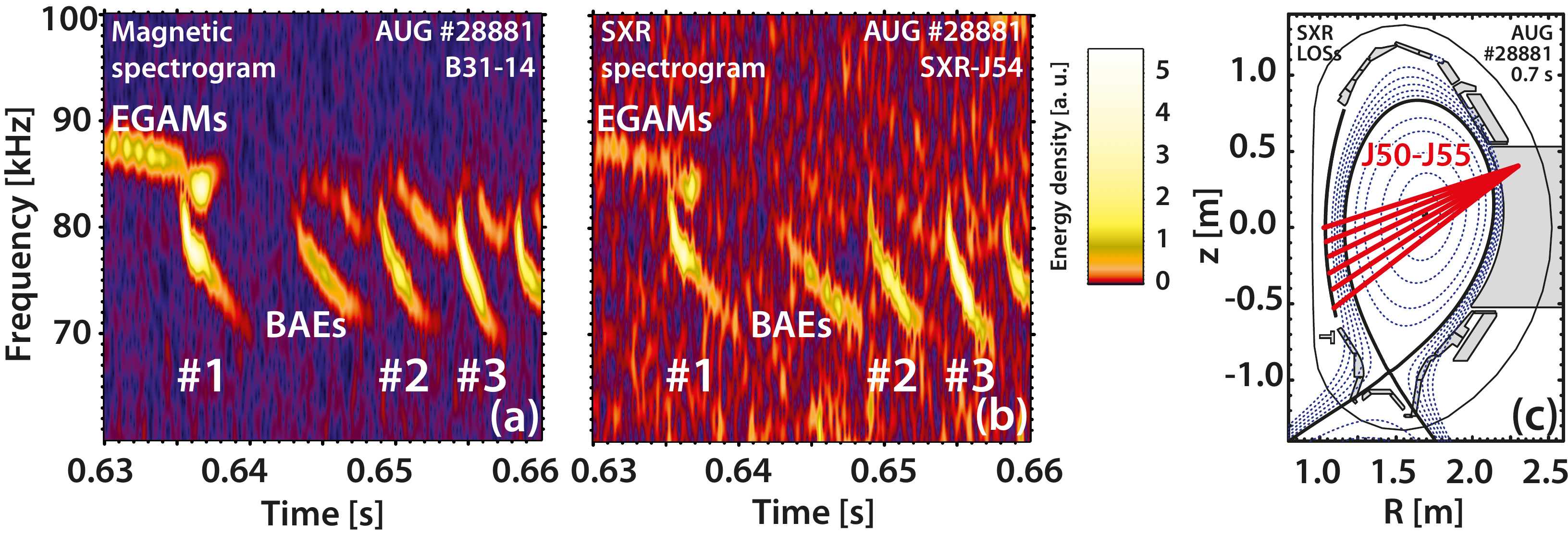}
  \caption{\textbf{(a)}~Chirping BAEs with decreasing frequency in the range of $70-85$~kHz are visible on the magnetic spectrogram. Modes around $90$~kHz are EGAMs (see later). Numbers denote the chirps which were investigated in detail. The width of the applied Gabor atom used to evaluate STFT is $\sigma_t = 0.125$~ms. \textbf{(b)}~Chirps are also visible on the SXR spectrogram ($\sigma_t = 0.125$~ms). \textbf{(c)}~The signal-to-noise ratio was appropriate for the radial structure analysis on 6 LOSs of SXR camera J: J50-J55.}
  \label{fig:bae_spectrogram}
\end{figure}

First, the results from the observation of beta-induced Alfv\'{e}n eigenmodes (BAEs) are presented.
As it is shown in table~\ref{tab:shots}, BAEs were observed in seven cases in this series of discharges.
However, the signal-to-noise ratio of the SXR signals was only appropriate for further analysis in discharge \#28881.
In this discharge (see figure~\ref{fig:28881} for the main plasma parameters) on-axis NBI heating was applied early (from $t = 0.35$ s), then off-axis NBI was used from $t = 0.6$ s.
Shortly after, bursting modes appeared at around 80 kHz as it is shown on the magnetic and SXR spectrograms in figure~\ref{fig:bae_spectrogram}a~and~b, respectively.

\begin{figure}[htb!]\centering
  \includegraphics[width = 145mm]{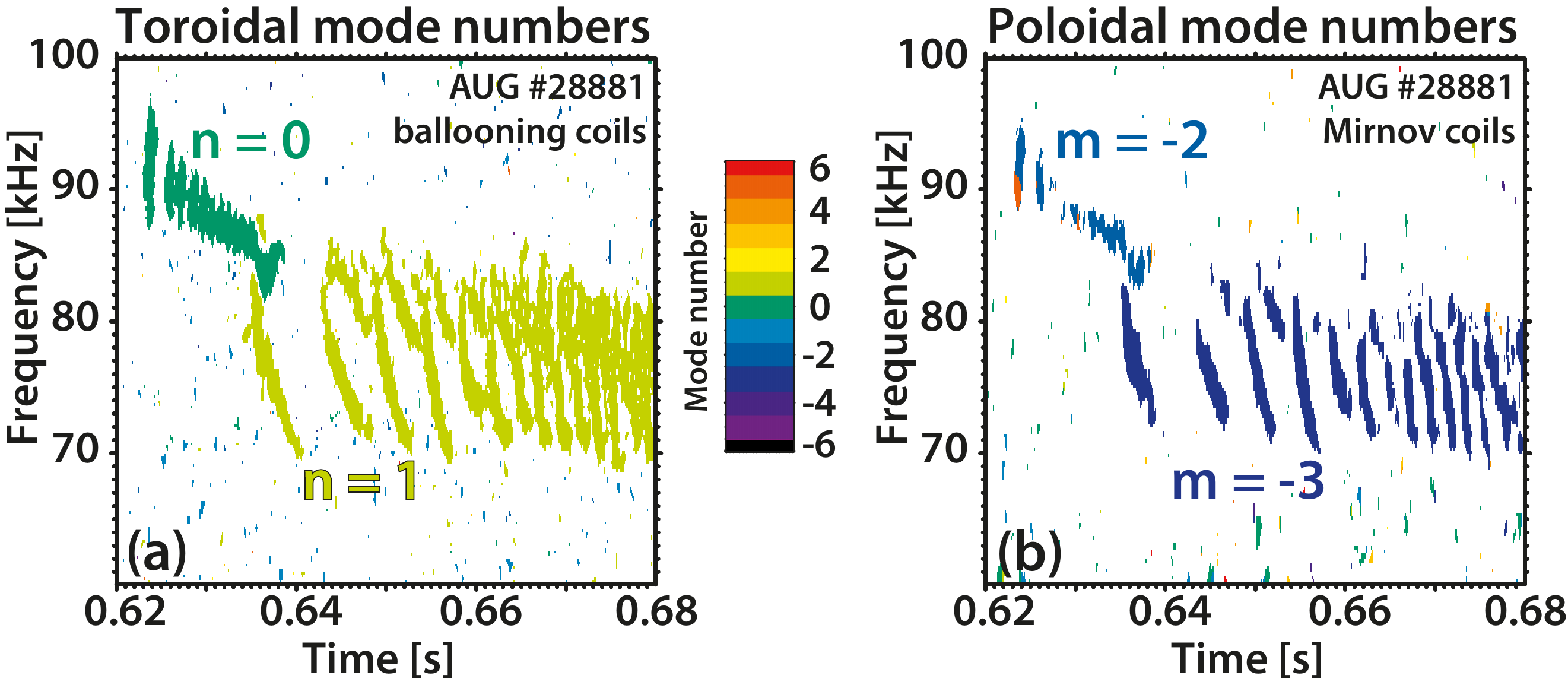}
  \caption{The result of time-frequency resolved mode number calculation. \textbf{(a)} The toroidal mode numbers are plotted only in time-frequency points where the residual of the fit is lower than 5~\% of the maximum. The positive sign means that the mode propagates counterclockwise in the lab frame. (The plasma current was counterclockwise and the toroidal magnetic field was clockwise.) \textbf{(b)} The poloidal mode numbers are plotted only in time-frequency points where the value of the minimum coherence is higher than~0.3. The negative sign means that the mode propagates in the ion diamagnetic drift direction in the lab frame.}
  \label{fig:bae_mode_number}
\end{figure}
An important step in mode identification is determining the toroidal and poloidal mode numbers.
The toroidal mode number was determined from the signals of the ballooning coils (see figure~\ref{fig:magnetic_probe_position}a), and the Mirnov-coils (see figure~\ref{fig:magnetic_probe_position}b) were used to evaluate the poloidal mode number.
The time-frequency resolved mode number analysis was carried out using an STFT based method~\cite{horvath15reducing}.
For toroidal mode numbers, the measured frequency dependent transfer functions of the ballooning coils~\cite{horvath15reducing} were taken into account which resulted in a clear improvement in the fitting quality, but did not change the value of the best fitting mode number.
The results are presented in figure~\ref{fig:bae_mode_number}.
In figure~\ref{fig:bae_mode_number}a the toroidal mode numbers are plotted only in time-frequency points where the residual of the fit is lower than 5~\% of the maximum.
This filter ensures that only well-fitting mode numbers are taken into account.
It is visible that modes above~85~kHz have $n=0$ toroidal mode number and the downchirping modes below 85 kHz have $n=1$ toroidal mode number.

The poloidal mode number calculation in figure~\ref{fig:bae_mode_number}b shows that the poloidal mode number of the modes above~85~kHz is $m=-2$ and the poloidal mode number of the downchirping modes below 85 kHz is $m=-3$.
In this case we had no information on the frequency dependent transfer function of the Mirnov coils which can lead to systematic errors in the mode number fitting.
Due to the large number of probes and low mode numbers, the lack of transfer functions did not lead to incorrect mode numbers.
However, because of the systematic errors the fitting residuals are increased, therefore in figure~\ref{fig:bae_mode_number}b a different filter method was required than in the case of toroidal mode numbers: the mode numbers are plotted only in time-frequency points where the value of the minimum coherence~\cite{pokol10wavelet} is higher than~0.3.
The averaging of the coherence calculation was performed with a moving boxcar kernel with
a width of $6 \sigma_t$, which is three times longer than the the width of the applied Gabor atom.
According to their mode numbers ($n=0$, $m=-2$) the bursting modes above 85 kHz are most probably EGAMs.

\begin{figure}[htb!]\centering
  \includegraphics[width = 80mm]{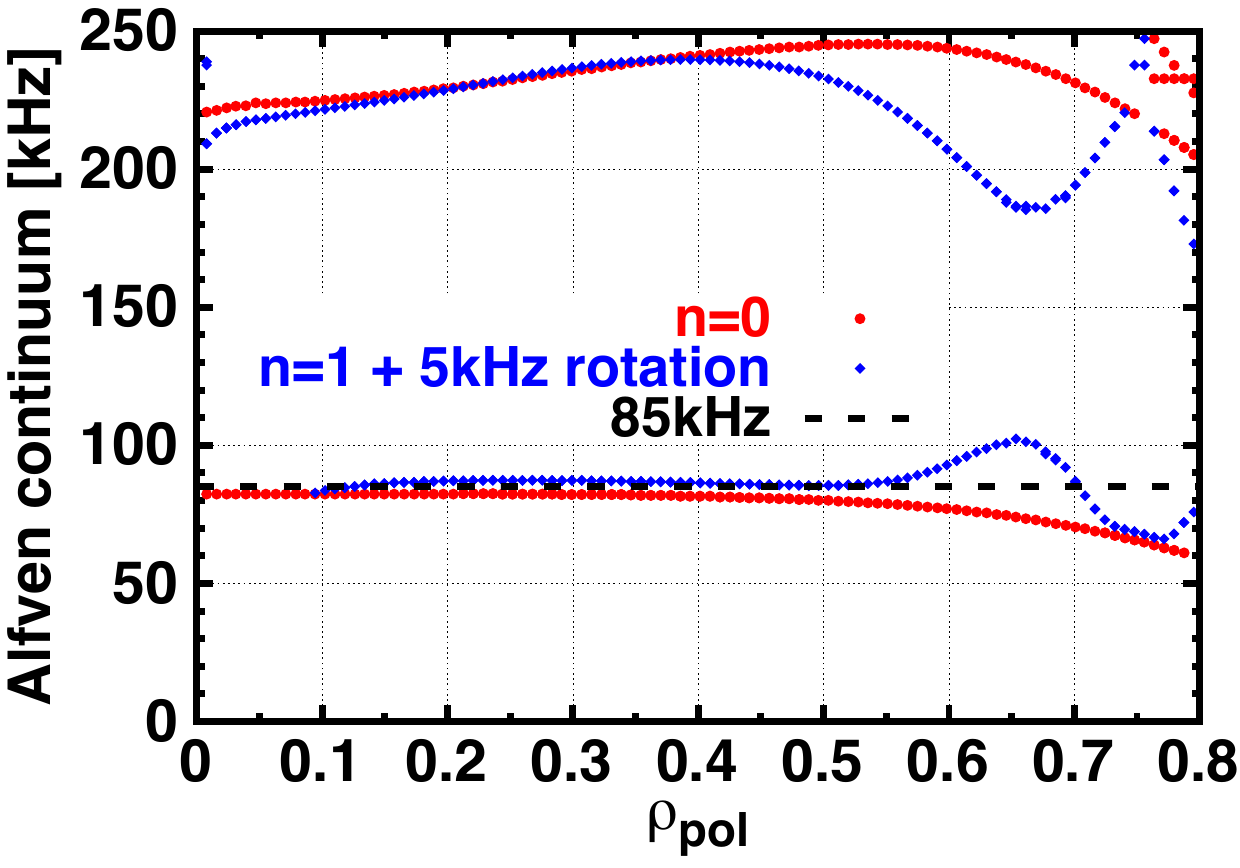}
  \caption{The kinetic spectrum (LIGKA~\cite{lauber07ligka}), including EP pressure) in \#28881 for $n=1$ ($q_0=3.1$, 5 kHz plasma rotation added) and $n=0$ ($q_0=3.1$).}
  \label{fig:28881_continuum}
\end{figure}

The downchirping $n=1$ mode is emerging from the BAE gap which can be demonstrated by analysing the kinetic spectrum in figure~\ref{fig:28881_continuum} (calculated using LIGKA~\cite{lauber07ligka}), including EP pressure) for $n=1$ ($q_0=3.1$, 5 kHz plasma rotation added) and $n=0$ ($q_0=3.1$).
One can conclude from the spectrum that the experimental mode chirping ($\lesssim 85$~kHz) starts below the continuum.
The fact that an $n=0$ mode is observed in the experiment just before the $n=1$ mode starts at the same frequency (just shifted by a small amount of toroidal rotation $\sim5$~kHz) is a direct proof of this interpretation.

The radial structure analysis of BAEs was carried out by using the LOSs of SXR camera J.
These LOSs are shown in figure~\ref{fig:bae_spectrogram}c with red lines.
Based on the signal-to-noise ratio of the SXR signals three chirps were selected for further analysis.
The soft X-ray spectrogram of channel J54 is shown in figure~\ref{fig:bae_spectrogram}b.
The instantaneous amplitude of the mode on each LOS is calculated by using the first order approximation defined in~\eqref{eq:stft_linear_4}.
The mode frequency was evaluated from the magnetic spectrogram, because the magnetic signals have a higher signal-to-noise ratio.

\begin{figure}[htb!]\centering
  \includegraphics[width = 150mm]{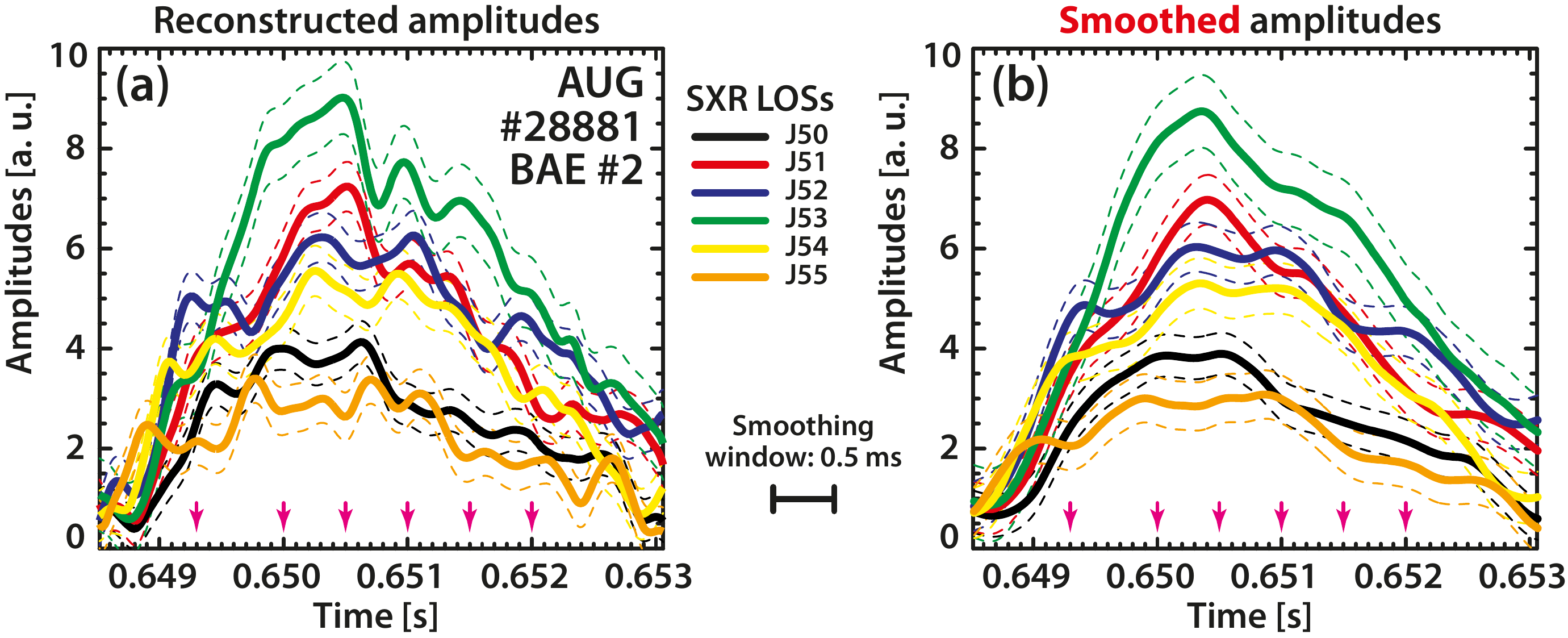}
  \caption{The time evolution of the oscillation amplitude of BAE chirp \#2 (see figure~\ref{fig:bae_spectrogram}) on the different SXR LOSs. \textbf{(a)}~Reconstructed amplitudes by using linear chirp approximation (eq.~\eqref{eq:stft_linear_4}). \textbf{(b)}~Amplitudes smoothed by a moving average with boxcar kernel of $0.5$~ms width.}
  \label{fig:bae_amps_maxorder1}
\end{figure}

\begin{figure}[htb!]\centering
  \includegraphics[width = 150mm]{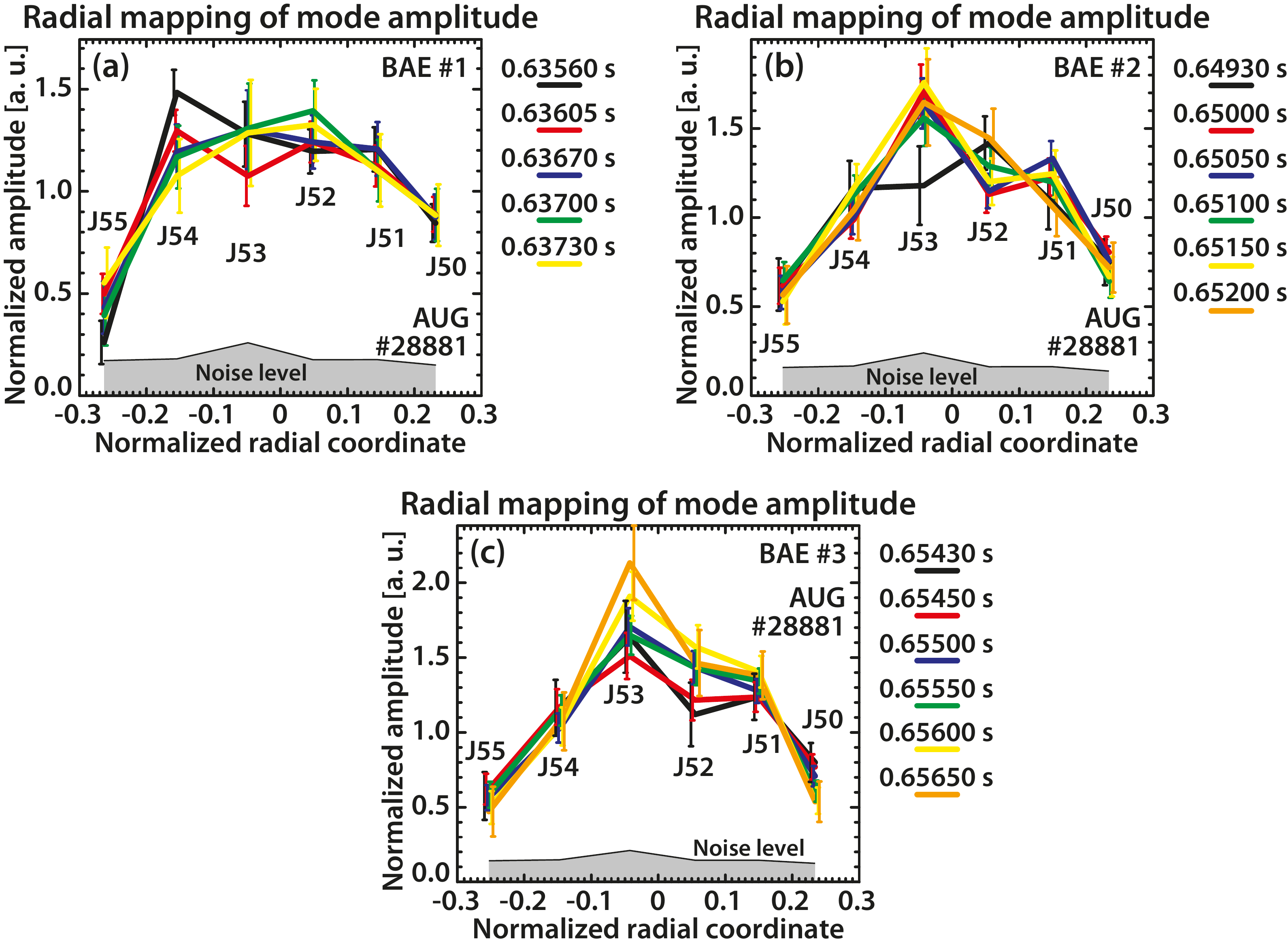}
  \caption{The radial mapping of the oscillation amplitude of BAEs. The results do not show significant changes in the radial distribution of the oscillation amplitudes, in agreement with present theoretical understanding.}
  \label{fig:bae_radmap}
\end{figure}

The time evolution of the instantaneous amplitude of chirp \#2 on the different SXR LOSs is shown in figure~\ref{fig:bae_amps_maxorder1}a.
In order to reduce the effect of the noise smoothing is applied by a moving average with boxcar kernel of $0.5$ ms width.
The smoothed amplitudes are presented in figure~\ref{fig:bae_amps_maxorder1}b.
To examine the changes in the radial structure, a radial mapping of the oscillation amplitudes was constructed.
This means that each LOS is labelled with the normalized poloidal flux ($\rho_{\mathrm{pol}}$) of the magnetic flux surface to which the LOS is tangential~\cite{gude14identification}.
To distinguish between the LOSs that pass above and below the magnetic axis, the radial coordinate of the latter is assigned with $-\rho_{\mathrm{pol}}$.
The oscillation amplitude was calculated for each LOS and it is plotted as the function of the radial coordinate (see figure~\ref{fig:bae_radmap}).
This way the radial mapping of the oscillation amplitudes can be evaluated at any time instant.
Since this work focuses on the relative changes in the radial structure of the modes, each curve of the radial map was normalized with its integral.
This radial map for chirp \#2 is shown in figure~\ref{fig:bae_radmap}b at time instances indicated with pink arrows in figure~\ref{fig:bae_amps_maxorder1}.
Similarly, figure~\ref{fig:bae_radmap}a and figure~\ref{fig:bae_radmap}c show the time evolution of the oscillation amplitude distribution for chirps \#1 and \#3.
Note that the x-axis in figure~\ref{fig:bae_radmap} is not a ``proper'' radial coordinate, it is the radial coordinate associated with the given SXR LOS.
Thus, the radial location of BAEs could not be deduced from the results in figure~\ref{fig:bae_radmap}.
According to the mode number analysis (see figure~\ref{fig:bae_mode_number}), the observed BAEs are expected to be localized at the $q=3$ surface.
The error bars in figure~\ref{fig:bae_amps_maxorder1} and \ref{fig:bae_radmap} are derived from the estimated background noise level of the SXR spectrograms in time-frequency regions showing no mode activity.
The results do not show significant and systematic changes in the radial distribution of the oscillation amplitudes.
This suggests that if there is any change in the radial structure it is smaller than the uncertainty of our measurement, which is $8-12$~\% at the maximum amplitude.
It also means that to detect changes in the radial structure, the oscillation amplitudes on the different LOSs should change (relative to each other) more than $8-12$~\% at the maximum oscillation amplitude.
Where the amplitude is smaller (the relative uncertainty is bigger), the detection threshold is higher.
The observation of the small change is consistent with the physical picture that BAEs are normal modes of the plasma in the BAE gap and the radial structure strongly depends on the background plasma parameters rather than on the EP distribution.

\section{Energetic particle driven GAMs}\label{sec:egam}

\begin{figure}[htb!]\centering
  \includegraphics[width = 80mm]{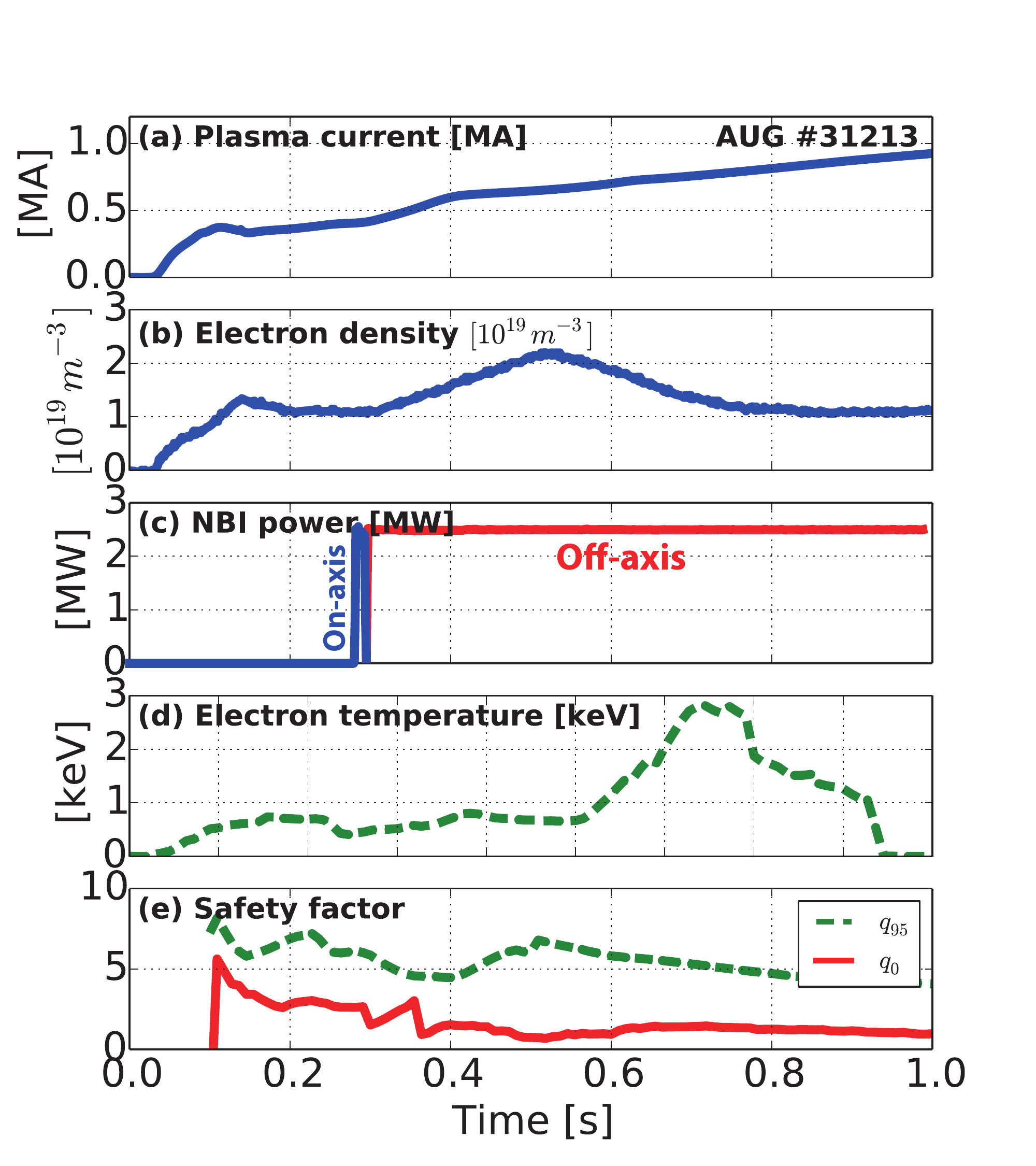}
  \caption{Main waveforms of discharge \#31213: \textbf{(a)}~plasma current, \textbf{(b)}~line-integrated core electron density, \textbf{(c)}~NBI power, \textbf{(d)}~Core electron temperature with green dashed line (no ion temperature measurement available) and \textbf{(e)}~safety factor $q$ at the magnetic axis with red solid line and $q_{95}$ with green dashed line.}
  \label{fig:31213}
\end{figure}

\begin{figure}[htb!]\centering
  \includegraphics[width = 130mm]{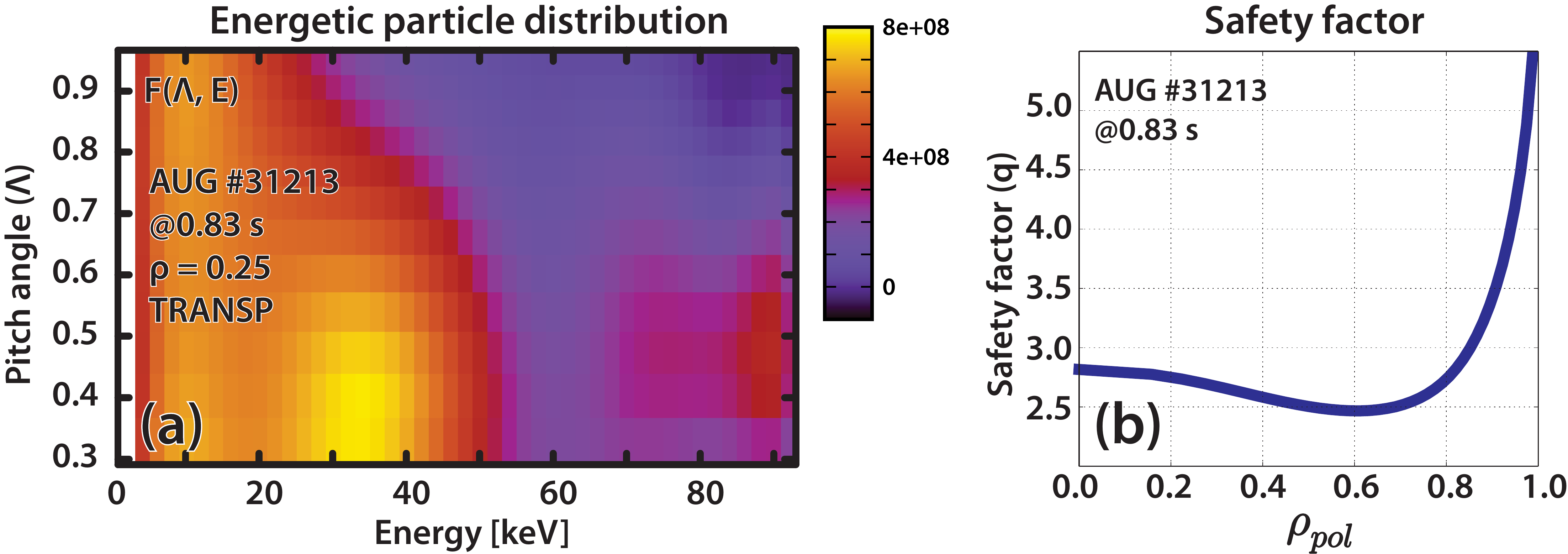}
  \caption{\textbf{(a)}~The fast distribution (calculated using the TRANSP code~\cite{pankin04tokamak}) and \textbf{(b)}~the q profile in discharge \#31213 at $t=0.83$ s, where the strongest EGAMs were observed.}
  \label{fig:egam_q}
\end{figure}

EGAMs were observed in the same (or similar) discharges (see table~\ref{tab:shots}), but in different time-frequency intervals as BAEs.
5 cases were found where the signal-to-noise ratio was appropriate for further analysis.
The strongest EGAMs were observed in discharge \#31213 (see figure~\ref{fig:31213} for the main plasma parameters), where the NBI deposition was most off-axis (the angle of the beam-line was $7.13^{\circ}$ with respect to the horizontal axis).
In this discharge the evaluated fast ion distribution had a peak at $\rho_{\textrm{pol}} \approx 0.45$ according to the fast ion D-alpha (FIDA) spectroscopy measurements~\cite{geiger12fast, lauber14offaxis}.
The TRANSP~\cite{pankin04tokamak} simulation of the fast ion distribution is shown in figure~\ref{fig:egam_q}, along with the $q$ profile at $0.83$~s when the strongest EGAMs appeared in discharge \#31213.

\begin{figure}[htb!]\centering
  \includegraphics[width = 145mm]{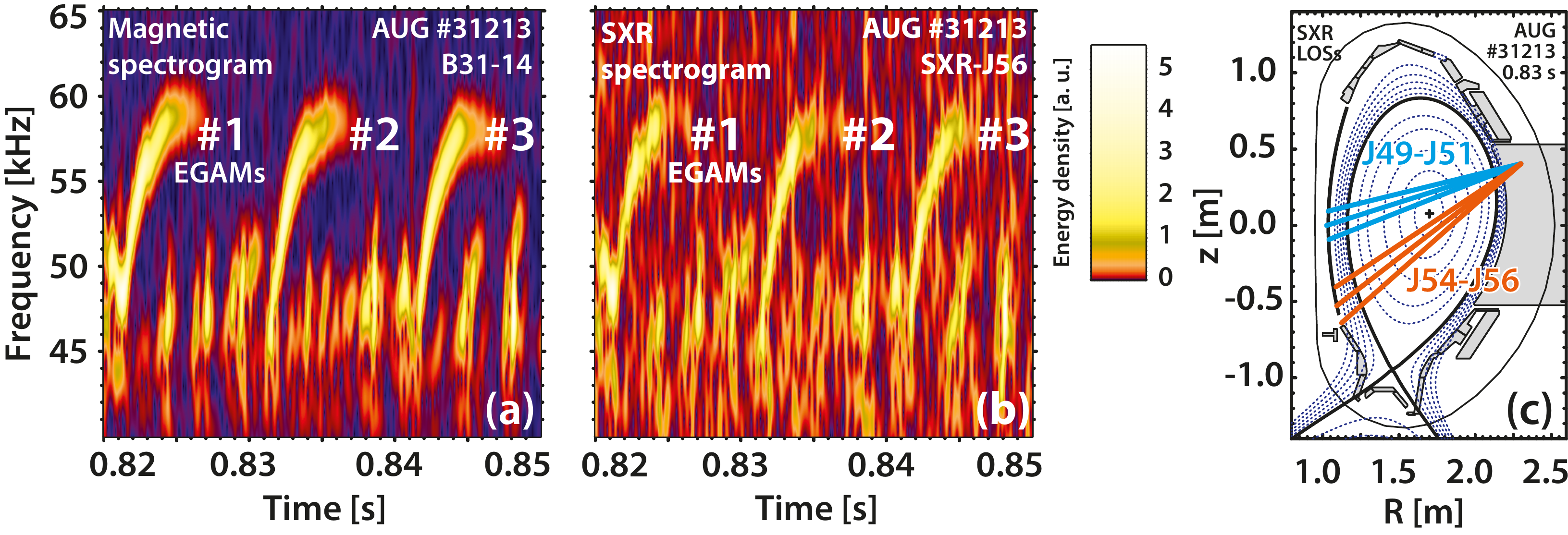}
  \caption{\textbf{(a)}~Chirping EGAMs with increasing frequency in the range of $45-60$ kHz are visible on the magnetic spectrogram.  The width of the applied Gabor atom used to evaluate STFT is $\sigma_t = 0.125$ ms. \textbf{(b)}~Chirps are also visible on the SXR spectrogram. Numbers denote the chirps which were investigated in detail ($\sigma_t = 0.125$ ms). \textbf{(c)}~The signal-to-noise ratio was appropriate for the radial structure analysis on 6 LOSs of SXR camera J: J49-51 \& J54-56.}
  \label{fig:egam_spectrogram}
\end{figure}

\begin{figure}[htb!]\centering
  \includegraphics[width = 145mm]{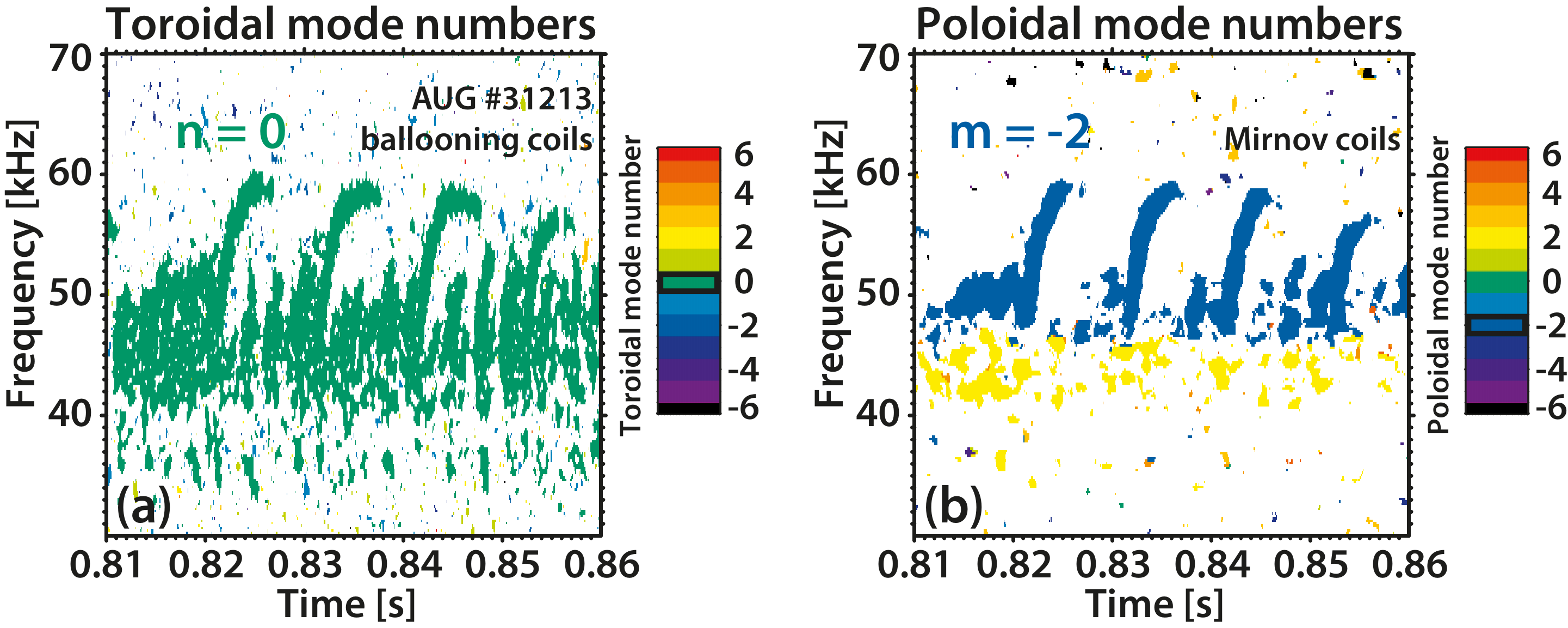}
  \caption{The result of time-frequency resolved mode number calculation. \textbf{(a)}~The toroidal mode numbers are plotted only in time-frequency points where the residual of the fit is lower than the 3~\% of the maximum. \textbf{(b)}~The poloidal mode numbers are plotted only in time-frequency points where the value of the minimum coherence is higher than~0.3.}
  \label{fig:egam_modenum}
\end{figure}

For mode identification and mode number analysis again the magnetic pick-up coils were used.
The time evolution of the mode frequency was traced using the same ridge-following algorithm as for BAEs.
The magnetic spectrogram from discharge \#31213 which shows the investigated 3 consecutive EGAMs is presented in figure~\ref{fig:egam_spectrogram}a.
The mode number analysis of EGAMs was carried out in a similar way as it was done for the BAEs.
The results of the mode number evaluation carried out on the magnetic signals are presented in figure~\ref{fig:egam_modenum}, where it is visible that all three EGAMs have $n=0$ toroidal and $m=-2$ poloidal mode number.
These mode numbers correspond to the sideband coupled mode.

The next issue to be resolved is to determine the radial location of the mode.
One estimation of the radial location can be given from the SXR measurements.
Starting from the plasma core, the first SXR LOS on which the mode is not visible gives an outer boundary of the mode location.
Considering the SXR spectrograms the investigated EGAMs are located inside the $\rho_{\mathrm{pol}}=0.4$ surface.
The radial location of the mode was also estimated from the frequency modulated continuous wave (FMCW) reflectrometry measurements.
Four channels (K, Ka, Q and V bands) were used with fixed frequencies (24, 33, 36 and 53 GHz correspondingly) from each the low and high field sides.
The corresponding density values are $0.71$, $1.35$, $1.61$ and $3.48\cdot10^{19}$ $1/$m${}^{3}$.
The detected modes are visible on the spectrograms of channels with frequencies 33 and 36 GHz.
According to the density profile (from Integrated Data Analysis - IDA~\cite{fischer2010integrated}) this means that the mode is located around $\rho_{\mathrm{pol}} \sim 0.25-0.45$.

The time evolution of the radial structure of EGAMs was investigated in the same way as it was done for BAEs.
Again, the LOSs of SXR camera J were chosen.
The selected 6 LOSs where the signal-to-noise ratio was appropriate are shown in figure~\ref{fig:egam_spectrogram}c in a poloidal cross-section of AUG.
The soft X-ray spectrogram of channel J54 with 3 consecutive EGAMs is shown in figure~\ref{fig:egam_spectrogram}b.

\begin{figure}[htb!]\centering
  \includegraphics[width = 150mm]{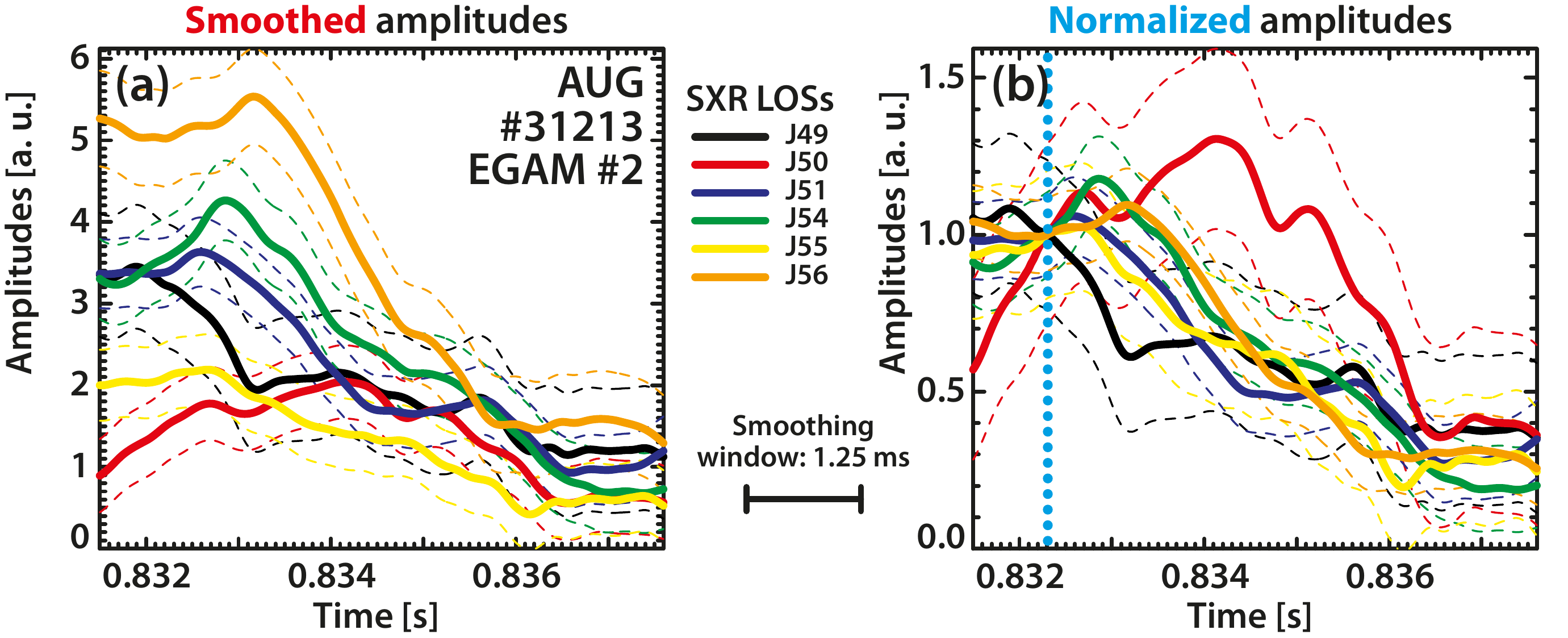}
  \caption{\textbf{(a)}~The time evolution of the oscillation amplitude of chirping EGAM \#2 (see figure~\ref{fig:egam_spectrogram}) on the different SXR LOSs. The amplitudes were smoothed with a moving boxcar kernel of $1.25$ ms width. \textbf{(b)}~The amplitudes shown in figure~\ref{fig:egam_amps_maxorder1}a are normalized to the amplitude at the beginning of the chirp (indicated with dotted cyan line) for further analysis.}
  \label{fig:egam_amps_maxorder1}
\end{figure}

The amplitude of the mode is evaluated for all LOSs by using the formula defined in~\eqref{eq:stft_linear_4}.
The reconstructed amplitudes of chirp \#2 (see figure~\ref{fig:egam_spectrogram}) as a function of time are presented in figure~\ref{fig:egam_amps_maxorder1}a for channels shown in figure~\ref{fig:egam_spectrogram}c.
In this figure, the amplitudes were smoothed by a moving average with boxcar kernel of $1.25$ ms width.
The uncertainty of the result is indicated by the dashed lines.

It is already visible in figure~\ref{fig:egam_amps_maxorder1} that the time evolution of the amplitude is different on the different channels.
Since the SXR measurement is not well localized, the exact changes in the radial structure cannot be tracked, but its behaviour can be qualitatively described.
The amplitudes shown in figure~\ref{fig:egam_amps_maxorder1}a are normalized to the amplitude at the beginning of the chirp for further analysis.
The result is shown in figure~\ref{fig:egam_amps_maxorder1}b.
In order to investigate the changes of the radial structure, the radial mapping of the the normalized amplitudes was carried out.

\begin{figure}[htb!]\centering
  \includegraphics[width = 150mm]{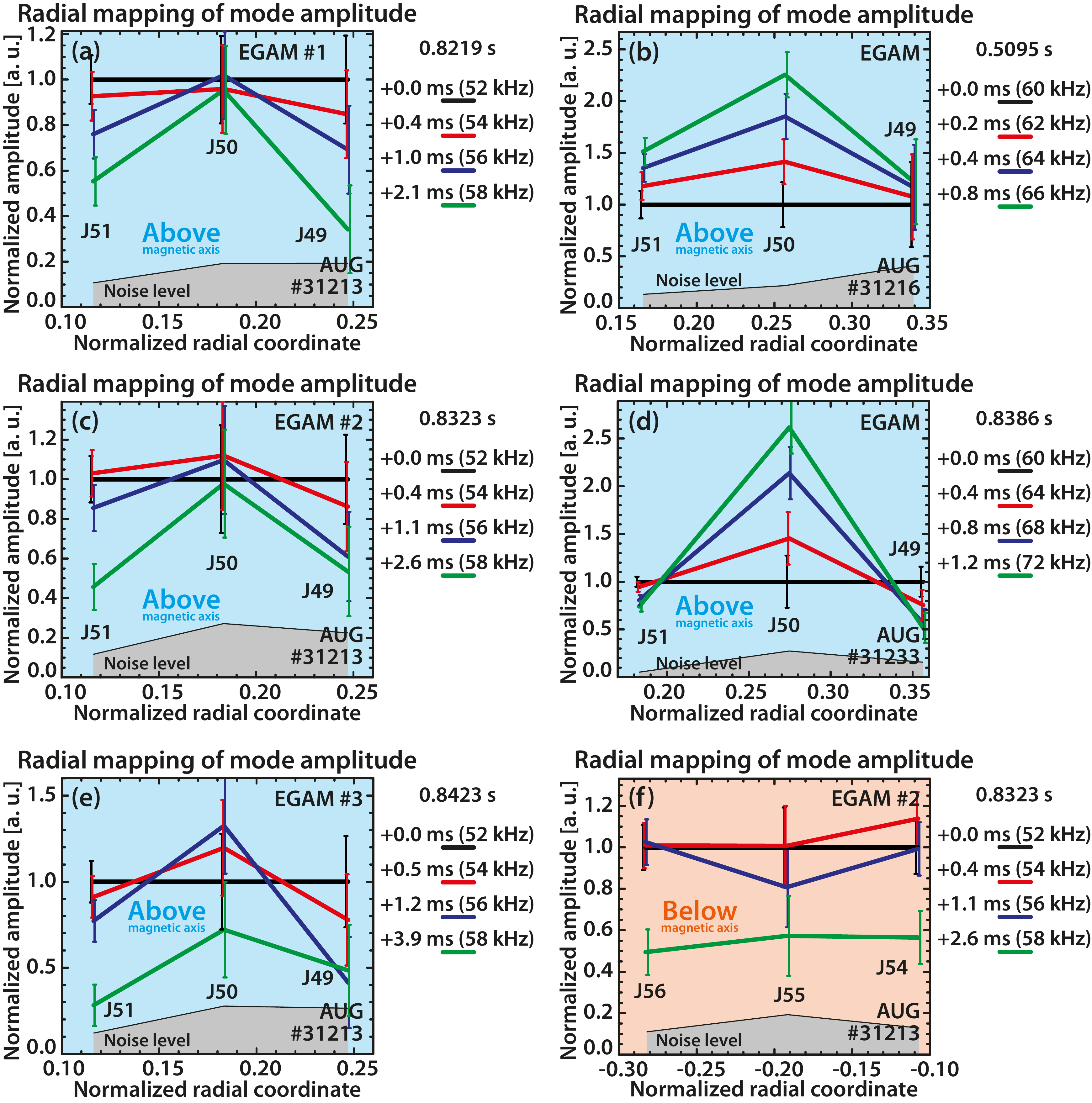}
  \caption{The radial mappings of the oscillation amplitude of EGAMs. \textbf{\mbox{(a-e)}}~5~cases showing the radial mapping from SXR LOSs located \emph{above} the magnetic axis. Shrinkage of the mode is visible in all cases. \textbf{(f)}~One case showing the radial mapping from SXR LOSs located \emph{below} the magnetic. Radial mappings calculated from channels pass below the magnetic axis do not show significant change.}
  \label{fig:egam_radmap}
\end{figure}

The LOSs that pass above and below the magnetic axis are handled separately and the time evolution of the radial mappings are shown in figure~\ref{fig:egam_radmap}.
On LOSs passing above the magnetic axis (figure~\ref{fig:egam_radmap}a-e), a shrinkage of the mode is visible, since as time evolves the relative amplitude in the middle channel is rising compared to the outer channels.
One example for the radial mapping calculated from channels that pass below the magnetic axis is presented in figure~\ref{fig:egam_radmap}f.
Regarding all 5 cases, radial mappings calculated from channels that pass below the magnetic axis do not show significant change.
This is most probably due to the longer distance from the observation point to the mode of these LOSs as it is visible in figure~\ref{fig:egam_spectrogram}c.
The LOSs passing below the magnetic axis observe a bigger volume, detecting more radiation from the background plasma and relatively less from the flux surfaces where the mode is.
However, the shrinkage of the mode is visible in radial mappings calculated from channels above the magnetic axis.
As it is shown in figure~\ref{fig:egam_radmap}a-e, this shrinkage was significant in all 5 investigated cases.

This observed shrinkage in the mode structure of EGAMs during the non-linear chirping phase is consistent with our present theoretical understanding.
The resonance condition to drive EGAMs shows that the $\omega$ mode frequency is equal to the $\omega_t$ transit frequency of the interacting ions:
\begin{equation}\label{eq:egam_resonance}
  \omega - \omega_t = 0.
\end{equation}

The exact phase space and real space coordinates of the wave-particle interaction are affected by several factors.
First, the EGAM drive is proportional to the velocity space gradient in the EP distribution function.
Second, the strength of the drive is proportional to the EP density.
In addition, the spatial dependence of the damping is an important factor.
From the SXR measurements, the radial position of the mode is expected to be at $\rho\approx0.25$, thus the EP distribution was investigated at this position.

\begin{figure}[htb!]\centering
  \includegraphics[width = 100mm]{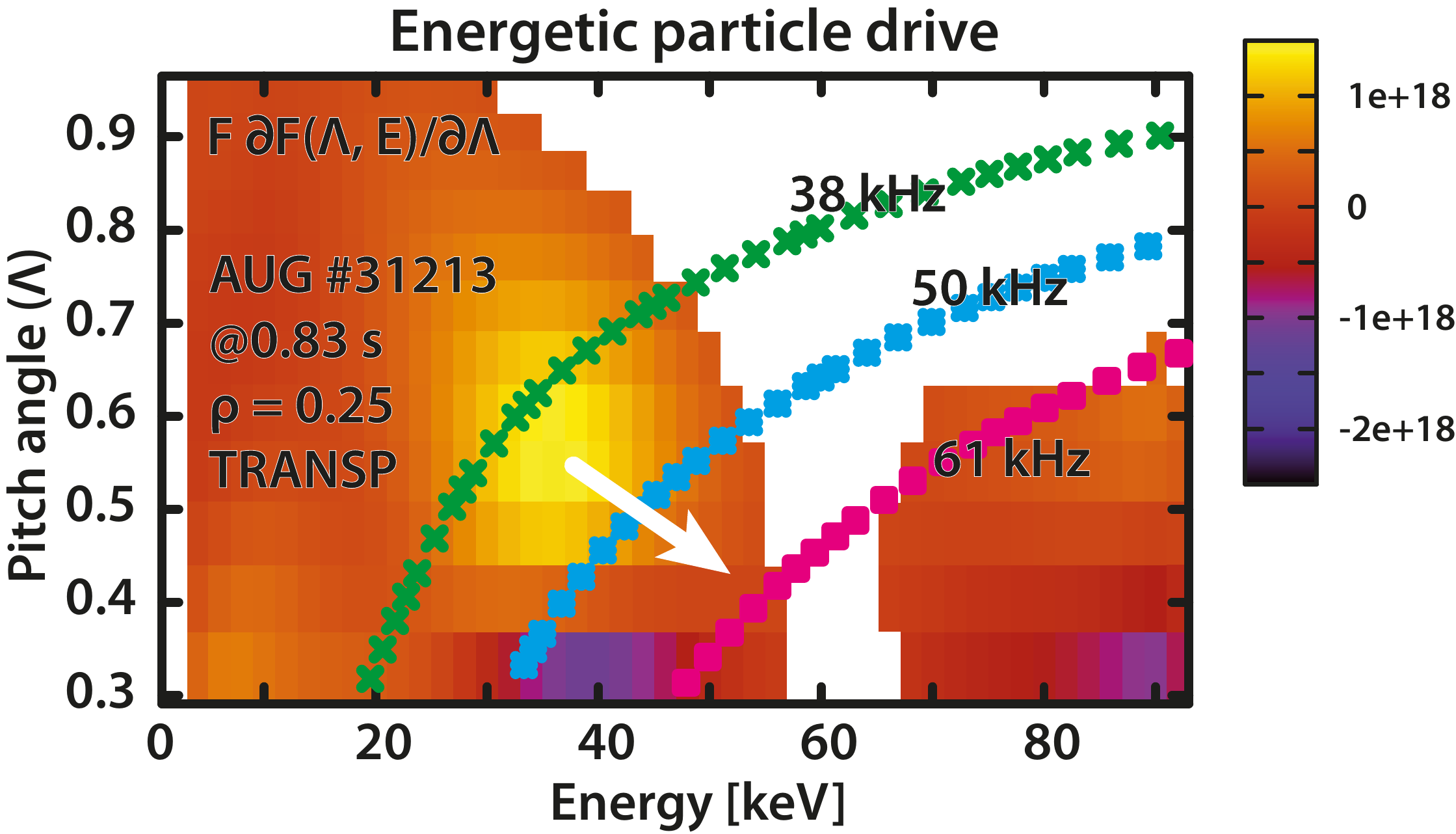}
  \caption{The contour plot shows the EP density ($F(\Lambda, E)$) times its derivative ($\partial F/\partial\Lambda$) as a function of the pitch angle ($\Lambda$) and energy ($E$) at $\rho\approx0.25$ in discharge \#31213 at $t=0.83$ s, where EGAMs were observed. The region where the fast ion distribution is close to zero is indicated with white. The coloured curves show the $(\Lambda, E)$ coordinates of particles with different transit frequencies (38, 50 and 61 kHz).}
  \label{fig:egam_resonance}
\end{figure}

The EP distribution at $\rho\approx0.25$ was calculated using the TRANSP code~\cite{pankin04tokamak} at the time instance when EGAMs appeared.
The EP density ($F(\Lambda, E)$) times its derivative ($\partial F/\partial\Lambda$) is plotted in figure~\ref{fig:egam_resonance} as a function of the pitch angle ($\Lambda = \mu B_0/E$) and energy ($E$).
The EP distribution can excite the mode when the drive overcomes the damping.
This can be reached where the drive is highest, i.e. at the peak of the $F\cdot\partial F/\partial\Lambda$ function which is around $(\Lambda = 0.75, E = 40$ keV$)$ in this case, as is shown in figure~\ref{fig:egam_resonance}.
The coordinates of the peak of the $F\cdot\partial F/\partial\Lambda$ function determine the parameters of interacting particles, which are passing particles.
The coloured curves in figure~\ref{fig:egam_resonance} show the $(\Lambda, E)$ coordinates of particles with different transit frequencies (38, 50 and 61 kHz).
The curves corresponding to higher frequencies move from the green curve ($38$ kHz) towards the pink curve ($61$ kHz) as it is shown by the white arrow.

\begin{figure}[htb!]\centering
  \includegraphics[width = 70mm]{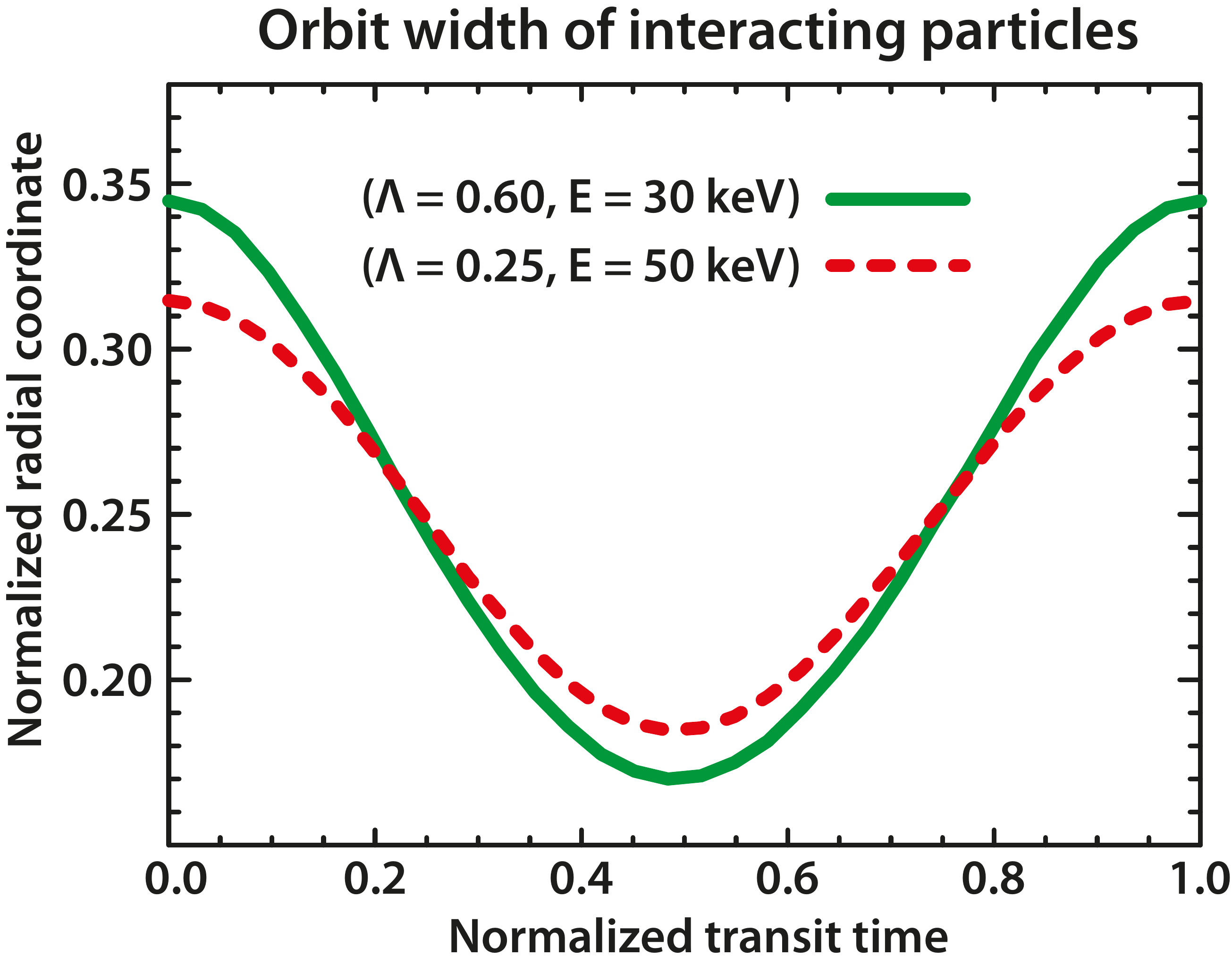}
  \caption{The orbit of EPs with different velocity space coordinates $(\Lambda, E)$. The radial coordinate of the particle is plotted as the function of the normalized circulation time.}
  \label{fig:egam_orbit}
\end{figure}

The mode frequency of the observed EGAMs starts at around $45$ kHz and increases until $60$ kHz as it is shown in figure~\ref{fig:egam_spectrogram}.
The curve corresponding to the initial $45$ kHz frequency would intersect the region of the maximum peak in figure~\ref{fig:egam_resonance}.
This means that - considering~\eqref{eq:egam_resonance} - the observed mode frequency is consistent with the simulated EP distribution function and the velocity space coordinates of the interacting particles are approximately $(\Lambda = 0.75, E = 40$ keV$)$.
The orbit of EPs with $\Lambda = 0.75$ pitch angle and $E = 40$~keV energy at $\rho\approx0.25$ is illustrated in figure~\ref{fig:egam_orbit} with solid green line where the radial coordinate of the particle is plotted as the function of the normalized circulation time.
It shows that these interacting particles have a $\Delta\rho\approx0.34-0.17=0.17$ orbit width.

The mode frequency increases up to $60$~kHz (see figure~\ref{fig:egam_spectrogram}).
This means that as time evolves, the resonance changes and shifts to particles with higher transit frequency.
The pink curve in figure~\ref{fig:egam_resonance} shows the phase space coordinates of particles which have $61$ kHz transit frequency.
During the time evolution of the mode, the resonance condition moves from $(\Lambda = 0.75, E = 40$~keV$)$ toward the pink curve, i.e. to the region of more passing particles (particles with smaller $\Lambda$).
The exact time evolution of the resonance condition cannot be determined from the equilibrium EP distribution, but most probably the resonance follows the ridge in the $F\cdot\partial F/\partial\Lambda$ function indicated with the white arrow in figure~\ref{fig:egam_resonance}.
More passing particles have a narrower orbit width which can explain the experimentally observed shrinkage of the radial mode structure since the mode structure of EGAMs is determined by the parameters of the drive, i.e. the orbit width of the interacting particles correlates to the mode extent.
The orbit of EPs with coordinates $(\rho = 0.25, \Lambda = 0.25, E = 50$~keV$)$ is shown in figure~\ref{fig:egam_orbit} with dashed red line.
It is shown in this figure that the orbit width corresponding to the end of the EGAM time evolution ($\Delta\rho\approx0.31-0.18=0.13$) is narrower with approximately 25~\% than the initial orbit width at the beginning of the interaction.
The consequence of the narrower orbit width is the shrinkage of the mode structure, which is consistent with the experimental observations.

\section{Summary and conclusions}\label{sec:conc}

The understanding of energetic particle (EP) driven plasma modes plays a key role regarding future burning plasma experiments.
Super-thermal EPs in tokamak plasmas can excite various instabilities, and the most important transport process of EPs in the plasma core is their interaction with these global plasma modes.
The non-linear behaviour of the mode amplitude and frequency may exhibit a wide range of different behaviours, which significantly influences the impact of the instabilities on the fast particle transport.
Therefore, in order to comprehensively understand the non-linear behaviour of EP-driven instabilities, the investigation of these modes is essential.
In this paper the rapid changes in the radial structure of beta induced Alfv\'{e}n eigenmodes (BAEs) and EP-driven geodesic acoustic modes (EGAMs) were experimentally investigated during the non-linear chirping phase.

For the extensive characterization of the modes three diagnostic systems were available, namely the magnetic pick-up coils, the soft X-ray (SXR) cameras and the reflectrometry measurements.
In order to deal with the transient behaviour of the phenomena, short-time Fourier transform was chosen as a basis of the data processing.
The time evolution of the radial structure of EP-driven modes was examined using the SXR measurements.

BAEs and EGAMs observed in the ramp-up phase of off-axis NBI heated plasmas in ASDEX Upgrade were investigated in details.
The radial structure of BAEs was unchanged within the uncertainty of the measurement.
This behaviour is consistent with the theory stating that the radial structure of BAEs -- as normal modes -- strongly depends on the background plasma parameters rather than on the EP distribution.

In the case of rapidly upward chirping EGAMs the analysis shows systematic shrinkage of the mode structure during the chirping phase.
This can be explained by the changing resonance condition in the velocity space of EPs.
The rising frequency of the mode indicates that, as time evolves, the EGAM is driven by more passing particles which have narrower orbit width.
Since the mode structure of EGAMs is sensitive to the EP distribution, the narrower orbit width of the interacting particles leads to the experimentally observed shrinkage of the mode structure.

\section*{Acknowledgments}

The authors would like to thank Matthias Willensdorfer for useful discussions on the ECE signals.
Authors at BME NTI acknowledge the support of Hungarian State grant NTP-TDK-14-0022.
This work has been carried out within the framework of the EUROfusion Consortium and has received funding from the Euratom research and training programme 2014-2018 under grant agreement No 633053.
The views and opinions expressed herein do not necessarily reflect those of the European Commission.
The main author acknowledges the support of FuseNet\footnote{\url{http://www.fusenet.eu}} -- the European Fusion Education Network -- within the framework of the EUROfusion Consortium.

\section*{References}
\addcontentsline{toc}{section}{References}
\bibliographystyle{unsrt}
\bibliography{references}

\end{document}